\documentclass[12pt]{article}

\usepackage[utf8]{inputenc}

\usepackage{csquotes}

\usepackage[style=alphabetic,
  sorting = nyt,
  sortcites=true,
  giveninits=true,
  date=year,
  isbn=false,
  maxbibnames=99,
  maxalphanames=4, 
  backend=biber]{biblatex}   
\DeclareFieldFormat*{titlecase}{\MakeSentenceCase{#1}}

\DeclareSourcemap{\maps[datatype=bibtex]{\map{\step[fieldset=shorthand, null]}}}

\DeclareSourcemap{
  \maps[datatype=bibtex]{
    \map[overwrite]{
      \step[fieldsource=doi, final]
      \step[fieldset=url, null]
      \step[fieldset=eprint, null]
    }  
  }
}

\DeclareSourcemap{
  \maps[datatype=bibtex, overwrite]{
    \map{
      \step[fieldset=language, null]
      \step[fieldset=month, null]
      \step[fieldset=urldate, null]
    }
  }
}

\AtEveryBibitem{%
  \clearlist{language}%
  \clearlist{month}%
  \clearlist{urldate}%
  \clearfield{urldate}%
}

\renewbibmacro{in:}{}

\DeclareFieldFormat[misc]{title}{\mkbibquote{#1}}
\DeclareFieldFormat[article,inproceedings]{volume}{\mkbibbold{#1}}
\DeclareFieldFormat[inproceedings]{series}{\mkbibitalic{#1}}

\usepackage[english]{babel}
 \usepackage{amsmath}
\usepackage{amsthm}
\usepackage{amssymb}
\usepackage{hyperref}
\usepackage{braket}
\usepackage{color}
\usepackage{comment}
\usepackage{graphicx}
\usepackage[margin=2.5cm]{geometry}
\usepackage[noblocks]{authblk}
\usepackage[noabbrev]{cleveref}

\usepackage{bbm}
\usepackage{enumerate}

\usepackage{todonotes}
\setuptodonotes{inline}

\usepackage{chngcntr}
\counterwithin{figure}{section}
\counterwithin{equation}{section}
\counterwithout{figure}{section}

\usepackage[UKenglish]{isodate}
\cleanlookdateon

\let\C\relax

\newcommand{\I}{\mathrm{i}}
 \newcommand{\R}{ \mathbb{R} }
  \newcommand{\Sph}{ \mathbb{S} }
\renewcommand{\S}{\mathbb{S}}
\newcommand{\C}{ \mathbb{C} }
\newcommand{\N}{ \mathbb{N} }
\newcommand{\Z}{ \mathbb{Z} }
\newcommand{\D}{\mathrm{d}}

\newcommand{\mcP}{\mathcal{P}}
\newcommand{\mcF}{\mathcal{F}}
\newcommand{\mcL}{\mathcal{L}}

\newcommand{\mfF}{\mathfrak{F}}

\newcommand{\mcV}{\mathcal{V}}
\newcommand{\mcW}{\mathcal{W}}

 \newcommand{\norm}[1]{\left\Vert #1 \right\Vert} 
 \newcommand{\abs}[1]{\left\vert #1 \right\vert}
\newcommand{\longip}[3]{\left\langle #1 \middle\vert #2 \middle\vert #3 \right\rangle}

\DeclareMathOperator{\diag}{diag}

\DeclareMathOperator{\tr}{tr}
\DeclareMathOperator{\Tr}{Tr}
\DeclareMathOperator{\sgn}{sgn}
\DeclareMathOperator{\spec}{spec}

\DeclareMathOperator{\tRe}{Re}

\newcommand{\ud}{\,\textnormal{d}}

\newcommand{\ee}{\mathrm{e}}

\newcommand{\ii}{\mathrm{i}}

\newcommand{\expect}[1]{\left\langle #1 \right\rangle}
\newcommand{\ip}[2]{\left\langle #1 \middle\vert #2 \right\rangle}

\usepackage{seqsplit}

\theoremstyle{plain}
\newtheorem{thm}{Theorem}[section]

\newtheorem{prop}[thm]{Proposition}

\newtheorem{lemma}[thm]{Lemma}

\theoremstyle{definition}

\newtheorem{rmk}[thm]{Remark}
\newtheorem{assumption}[thm]{Assumption}

\newtheorem{ex}[thm]{Example}
\newtheorem{remark}[thm]{Remark}

\crefname{thm}{theorem}{theorems}
\crefname{problem}{problem}{problems}
\crefname{lemma}{lemma}{lemmas}
\crefname{lem}{lemma}{lemmas}
\crefname{cor}{corollary}{corollaries}
\crefname{prop}{proposition}{propositions}
\crefname{conj}{conjecture}{conjectures}
\crefname{defn}{definition}{definitions}
\crefname{defi}{definition}{definitions}
\crefname{note}{note}{notes}
\crefname{ex}{example}{examples}
\crefname{remark}{remark}{remarks}
\crefname{rmk}{remark}{remarks}
\crefname{notation}{notation}{notations}
\crefname{assumption}{assumption}{assumptions}
\crefname{claim}{claim}{claims}
\crefname{claim*}{claim}{claims}

\allowdisplaybreaks

\title{Multi-band superconductors have enhanced critical temperatures}

\author[1]{Joscha Henheik\thanks{\href{mailto:joscha.henheik@ist.ac.at}{\nolinkurl{joscha.henheik@ist.ac.at}}}\,}
\author[2]{Edwin Langmann\thanks{\href{mailto:langmann@kth.se}{\nolinkurl{langmann@kth.se}}}\,}
\author[1]{Asbjørn Bækgaard Lauritsen\thanks{\href{mailto:alaurits@ist.ac.at}{\nolinkurl{alaurits@ist.ac.at}}}\,}

\affil[1]{Institute of Science and Technology Austria, Am Campus 1, 3400 Klosterneuburg, Austria.}
\affil[2]{Department of Physics, KTH Royal Institute of Technology, 106 91 Stockholm, Sweden.}

\date{\today}

\begin{document}
\maketitle

\begin{abstract}
We introduce a multi-band BCS free energy functional and prove that for a multi-band superconductor the effect of inter-band coupling can only increase 
the critical temperature, irrespective of its attractive or repulsive nature {and its strength}.
Further, for weak coupling and weaker inter-band coupling, 
we prove that the dependence of the increase in critical temperature on the inter-band coupling is 
(1) linear, if there are two or more {equally strongly superconducting} bands, or (2) quadratic, if there is only one dominating band. 
~
\end{abstract}

\section{Introduction and main results}
Shortly after the development of the celebrated Bardeen-Cooper-Schrieffer (BCS) theory of superconductivity \cite{bcs.original}, Suhl, Matthias and Walker \cite{Suhl.Matthias.ea.1959}, and independently Moskalenko \cite{moskalenko1959superconductivity}, introduced an extension of BCS theory allowing for more complex electronic band structures. These models for \emph{multi-band superconductors} were subsequently theoretically studied, e.g.~by Kondo \cite{kondo1963superconductivity} and Leggett \cite{leggett1966number} in the 1960s. 

Despite these quite early modifications of BCS theory, it took around two decades until the first experimental realization \cite{binnig1980two} of multi-band superconductivity in $\mathrm{Nb}$ doped $\mathrm{SrTiO}_3$. 
Still,
(probably) due to the relatively low critical temperature $T_c$ (below which the material becomes superconducting --- see \eqref{eq:Tc} below for a mathematical definition), {interest in multi-band superconductivity remained small for another two decades.} 
{This changed with proposals that high-temperature superconductivity in cuprates \cite{bednorz1986possible} exhibit multi-band structure \cite{kresin1990multigap, muller1995possible}.} 
 The most flourishing period of research on multi-band superconductivity was kicked off by the discovery \cite{nagamatsu2001superconductivity} of a relatively high $T_c \approx 39\mathrm{K}$ in the conventional superconductor $\mathrm{MgB}_2$, whose characteristic feature is the interaction of two different electronic bands \cite{souma2003origin}. After $\text{MgB}_2$, other materials, such as $\mathrm{NbSe}_2$ \cite{huang2007experimental}, or iron based 
{high-temperature superconductors} \cite{richard2011fe}, were found to be multi-band superconductors. 

One of the most important features of multi-band superconductors is a strong $T_c$-enhancing effect of the inter-band interactions. 
This was already pointed out by Kondo \cite{kondo1963superconductivity} and further studied in the physics literature 
\cite{bussmann2004enhancements, bussmann2011isotope, bussmann2017multigap, bussmann2019multi}. 
In particular either linear \cite{bussmann2017multigap} or quadratic \cite{bussmann2011isotope,bussmann2019multi,bussmann2004enhancements} $T_c$-enhancements 
are expected depending on the multi-band superconductor. 
The aim of this paper is to put these predictions on rigorous ground. 
We restrict ourselves to continuum models where the Brillouin zone is $\R^d$; 
we believe that 
{our methods would be applicable also to models with other Brillouin zones,}
see \Cref{rem:BZ} for further details.
We also mention that the results in the present paper are restricted to the leading order in the weak coupling parameter $\lambda$. It would be interesting to extend our results to higher order, in generalization of such results in the one-band case \cite{Hainzl.Seiringer.2008, langmann.et.al.2019, langmann2023universal}, but this is beyond the scope of the present paper.

\normalcolor 

Multi-band BCS theory has not been studied much in the mathematical physics literature 
with the exception being the work of Yang \cite{yang2005mathematical}.
In \cite{yang2005mathematical}, however, rather restrictive assumptions on the interaction are imposed.
In this paper, we give a much more general mathematical formulation 
of multi-band BCS theory, similarly to the single-band setting by Hainzl, Seiringer and others \cite{hainzl.hamza.seiringer.solovej, hainzl.seiringer.16} and study the effect of multi-band interactions on the critical temperature of the system. As our main results, we prove that: 

\begin{description}
\item[\Cref{thm.main.kappa.c}.] Inter-band couplings can only increase the critical temperature $T_c$, irrespective of its attractive or repulsive nature and its strength.
\item[\Cref{thm:main}.] For weak coupling and weaker inter-band coupling, $T_c$ depends either (1) linearly or (2) quadratically on the inter-band coupling for (1) two or more equally strongly superconducting bands or (2) a single dominating band. 
\end{description}

\noindent \textit{Structure of the paper.}
In Section \ref{subsec:multiband}, we provide the mathematical formulation of multi-band BCS theory. 
Afterwards, in Section \ref{subsec:results}, we formulate our main results, whose proofs are given in Section~\ref{sec:proof}. 
In \Cref{app:derivation} we give a heuristic derivation of the 
multi-band BCS functional \eqref{eq:mbBCS} from a many-body Hamiltonian on a physics level of rigor, 
and in \Cref{sec.notation} we collect some basic notation used in the paper.

\subsection{Multi-band functional, gap equation, and critical temperature} \label{subsec:multiband}
We consider a gas of fermions in $\R^d$ for $d = 1,2,3$ at temperature $T > 0$. 
The particles are assumed to occupy $n \in \N$ \emph{bands} (alternatively, they come in $n$ different species), characterized by different dispersion relations $\epsilon_{a}(p)$ (i.e.~a relation between momentum $p$ and energy $\epsilon$) for $a = 1, ... , n$, which we assume to satisfy the following (cf.~\cite{hainzl.seiringer.2010}). 
\begin{assumption}[On the dispersion relation] \label{ass:dispersion}
For every $a \in \{1,...,n\}$, we have that the zero set of $\epsilon_{a}(p)$ is a manifold (a \emph{generalized Fermi surface}) 
\begin{equation} \label{eq:Sa}
S_a := \{ p \in \R^d : \epsilon_{a}(p) = 0 \} \subset \R^d
\end{equation}
of codimension one, which is not necessarily connected but consists of finitely many components. Moreover, there exists some $\sigma > 0$ and a compact neighborhood $\Omega \subset \R^d$ of $S_a$ containing $S_a$, such that $\mathrm{dist}(S_a, \Omega^c) \ge \sigma$ {($\Omega^c$ is the complement of $\Omega$ in $\R^d$)}. For $\epsilon_{a}$ we further assume that 
\begin{itemize}
\item[(i)] it is locally bounded, measurable, reflection-symmetric ($\epsilon_a(-p) = \epsilon_a(p)$) and satisfies $\epsilon_a \in C^2(\Omega)$; 
\item[(ii)] its gradient $\nabla \epsilon_a(p)$ does not vanish in $\Omega$; 
\item[(iii)] there exist constants $c, C > 0$ such that $\epsilon_{a}(p) \ge c p^2 - C$ for all  $a \in \{1,...,n\}$. 
\end{itemize}
\end{assumption}
Here and in the following we use the convention that $C$ denotes any positive constant and its value may change line by line.

Assumption \ref{ass:dispersion} is satisfied by all relevant (non-relativistic) dispersion relations, in particular the \emph{Sommerfeld band dispersion relation}
\begin{equation} \label{eq:effective}
	\epsilon_{a}(p) := \frac{p^2}{2 m_a} - \mu_a 
\end{equation}
with \emph{effective mass} $m_a > 0$ and \emph{effective chemical potential} $\mu_a > 0$. 
In previous works on superconductivity in the single-band case (see, e.g., \cite{hainzl.hamza.seiringer.solovej}), 
the authors always (with the exception of \cite{hainzl.seiringer.2010}) chose \eqref{eq:effective} with mass being set to $1/2$ by simple scaling.  
However, since, even if we restrict to Sommerfeld dispersion relations of the form \eqref{eq:effective} only, 
effective masses and effective chemical potentials can and will be different in different bands, this cannot be achieved in general in our multiband setting. 
Therefore, we keep the most general form of $\epsilon_{a}$ as specified in Assumption \ref{ass:dispersion}. This also allows for non-spherical Fermi surfaces $S_a$. 
In the physics literature, many multi-band models arise from one-band models with a non-spherical Fermi surface, and the Fermi surfaces in the emerging (multiple) bands are then also possibly non-spherical; see e.g.~\cite{markowitz1963effect, allen1976fermi, langmann1992fermi} (in the context of the Eliashberg theory).

The interaction between fermions in bands $a$ and $b$ is described by a two-body potential $V_{ab}$, for which we assume the following.
\begin{assumption}[On the interaction potential] \label{ass:basic}
	For any $a,b \in \{1,\ldots,n\}$ the interaction $V_{ab} = V_{ba} \in L^1(\R^d) \cap L^{p_V}(\R^d)$ is real-valued and reflection-symmetric (meaning $V_{ab}(-x) = V_{ab}(x)$)
	with $p_V =1$ if $d = 1$, $p_V \in (1, \infty)$ if $d = 2$, or $p_V = 3/2$ if $d = 3$. 
\end{assumption}

{We stress that the class of models we consider is large. It includes multi-orbital models obtained from a one-band model where the dispersion relation is rotation invariant but the two-body interaction potential is not; in such a case, $a$ can be identified with the angular momentum $\ell$ (which is $0,2,\ldots$ for $s$-, $d$-wave, and higher even angular momenta, respectively), and $V_{ab}$ for $a\neq b$ are interactions between different angular momentum channels. Another example are multiband models with Sommerfeld dispersion relations and rotation invariant interactions within and inbetween different bands; see Example~\ref{ex:expl} for details.} 

A multi-band BCS state $\Gamma$ is given by $n$ pairs of functions $(\gamma_a, \alpha_a)_{a=1}^n$ and can be conveniently represented as a $2n \times 2n$ matrix valued Fourier multiplier on $L^2(\R^d; \C^n) \oplus L^2(\R^d; \C^n)$ of the form
\begin{equation} \label{eq:Gamma}
	\hat{\Gamma}(p) = \begin{pmatrix}
	\hat{\gamma}(p) & \hat{\alpha}(p)\\ \overline{\hat{\alpha}(p)} & \mathbbm{1} - \hat{\gamma}(p)
\end{pmatrix},
\qquad 
\gamma = \diag[\gamma_{a}]_{a=1}^n,
\qquad \alpha = \diag[\alpha_{a}]_{a=1}^n
\end{equation}
for all $p \in \R^d$ {(the bar indicates complex conjutation)}. Here, for every band $a = 1, ... ,n$, $\hat{\gamma}_a(p)$ denotes the Fourier transform of the one particle density matrix and $\hat{\alpha}_a(p)$ is the Fourier transform of the Cooper pair wave function, both in band $a$. We require reflection symmetry of $\hat{\alpha}_a$, i.e.~$\hat{\alpha}_a(-p) = \hat{\alpha}_a(p)$, as well as $0 \le \hat{\Gamma}(p) \le 1$ as a matrix. 
In this paper, we study the following multi-band version of the standard BCS free energy functional 
\cite{leggett.diatomic,hainzl.hamza.seiringer.solovej}
(see also \cite{hainzl.seiringer.review.08,hainzl.seiringer.16,Hainzl.Seiringer.2008,hainzl.seiringer.scat.length,Henheik.Lauritsen.2022,Henheik.2022,lauritsen.energy.gap.2021}), 
which we will derive from a many-body Hamiltonian \cite{Suhl.Matthias.ea.1959, moskalenko1959superconductivity, kondo1963superconductivity, leggett1966number}  in Appendix \ref{app:derivation} on a physics level of rigor. It is given by
\begin{equation} \label{eq:mbBCS}
	\mathcal{F}_T[\Gamma] := \int_{\R^d} \sum_{a=1}^n \epsilon_{a}(p) \hat{\gamma}_{a}(p) \D p - T S[\Gamma] 
	+ \int_{\R^d} \sum_{a,b=1}^n V_{ab}(x) \overline{\alpha_a(x)} \alpha_b(x) \D x\,,
\end{equation}
where entropy per unit volume is defined as 
\begin{equation}\label{eqn.entropy.unit.vol.def}
	S[\Gamma] = - \int_{\R^d} \mathrm{Tr}_{\C^{2n}} \left[\hat{\Gamma}(p) \log \hat{\Gamma}(p)\right] \D p\,. 
\end{equation}

The variational problem associated with the BCS functional \eqref{eq:mbBCS} is studied on 
\begin{equation*}
	\mathcal{D} := \left\{   \Gamma \text{ as in } \eqref{eq:Gamma} : 0 \le \hat{\Gamma} \le 1, \ \hat{\gamma}_a \in L^1(\R^d, (1 + p^2) \D p), \ 
  \alpha_a \in H^1_{\rm sym}(\R^d,\!\ud x), \ a = 1, ... , n \right\}.
\end{equation*}
Here $H^1_{\textnormal{sym}}$ denotes the set of reflection-symmetric $H^1$-functions.
The following proposition, whose proof is completely analogous to those in \cite{hainzl.hamza.seiringer.solovej, hainzl.seiringer.16}, and so omitted, provides the foundation for studying this problem. 
\begin{prop} \label{prop:exofmin}
	Under Assumption \ref{ass:basic} on $V$, the BCS free energy is bounded below on $\mathcal{D}$ and attains its minimum. 
\end{prop}
The associated Euler-Lagrange equation is easily found to be
\begin{equation} \label{eq:EulerLagrange}
	(K_T^\Delta + V)\alpha = 0,
	\qquad K_T^\Delta = \diag\left[K_{T,a}^{\Delta_a}\right]_{a=1}^n
\end{equation}
where 
\begin{equation*}
	K_{T,a}^{\Delta_a}(p) = \frac{\sqrt{\epsilon_a(p)^2 + |{\Delta_a}(p)|^2}}{\tanh\left(\frac{\sqrt{\epsilon_a(p)^2 + |{\Delta_a}(p)|^2}}{2T}\right)} \,. 
\end{equation*}
Here, 
$V = (V_{ab})_{a,b=1}^n$ is the matrix of interactions, 
and we denoted the vector of gaps by $\Delta(p) = - 2\,  (2 \pi)^{-d/2} (\hat{V}\star \hat{\alpha})(p)$ 
with 
$(\hat V \star \hat \alpha)(p) := \int_{\R^d} \hat V(p-q) \hat\alpha(q) \D q$ the convolution.
{Further, we define $E_a(p) = \sqrt{\epsilon_a(p)^2 + |{\Delta_a}(p)|^2}$,
the modified dispersion relation(s) arising from the gap function(s) $\Delta_a$.}

An equivalent form of \eqref{eq:EulerLagrange} is the following natural analog 
 of the celebrated (standard single-band, see \cite{hainzl.hamza.seiringer.solovej}) \emph{BCS gap equation}, given by 
\begin{equation} \label{eq:GAP}
	\Delta_a(p) = -\frac{1}{(2\pi)^{d/2}} \sum_{b=1}^n \int_{\R^d} \hat V_{ab}(p-q) 
	\frac{\tanh\left(\frac{E_{b}(q)}{2T}\right)}{E_{b}(q)}
	\Delta_b(q) \ud q\,,
  \qquad a=1,\ldots,n.
\end{equation}
Written without the indices, the gap equation takes the following form, where the relevant objects are matrix-valued, 
$\Delta(p) = - (2\pi)^{-d/2}\int_{\R^d} \hat V(p-q) K_T^{\Delta}(q)^{-1} \Delta(q) \ud q $.

The system described by the functional $\mathcal{F}_T$ is \emph{superconducting} if and only if any minimizer $\Gamma$ of $\mathcal{F}_T$ has a non-vanishing vector of off-diagonal entries, $\alpha \not\equiv 0$ (or, equivalently, \eqref{eq:GAP} has a solution $\Delta \not\equiv 0$). The (a priori highly non-linear) question, whether a system is superconducting or not can be reduced to a \emph{linear} criterion involving the matrix-valued pseudo-differential operator with symbol $K_T(p) \equiv K_T^0(p)$. In fact, as can be shown completely analogously to \cite{hainzl.hamza.seiringer.solovej}, 
the system is superconducting if and only if the (matrix-valued) operator $K_T + V$ has at least one negative eigenvalue. 
Moreover, there exists a unique \emph{critical temperature} $T_c \ge 0$ being defined as 
\begin{equation} \label{eq:Tc}
	\boxed{T_c := \inf\{ T > 0 : K_T + V \ge 0 \}}\,,
\end{equation}
for which $K_{T_c} + V \ge 0$ and $\inf \mathrm{spec} (K_T + V) < 0$ for all $T < T_c$. It can easily be seen that, by Assumption \ref{ass:basic} and the asymptotic behavior $K_{T_c}(p) \gtrsim p^2$ for $|p| \to \infty$, the critical temperature is well-defined by invoking Sobolev's inequality \cite[Thm.~8.3]{analysis}. 

\subsection{Main results} \label{subsec:results}
In this paper, we study the effect of the interband coupling due to $V_{ab}$ for $a \neq b$ on the critical temperature $T_c$. More concretely, we rescale the original interaction matrix $V$ as
\begin{equation} \label{eq:Vscaling}
V \to \lambda V^{\mathrm{d}} + \kappa  \lambda V^{\mathrm{od}} \qquad \text{with} \qquad 
\lambda > 0
\quad \text{and} \quad \kappa \in \R\,, 
\end{equation}
where $V^{\mathrm{d}}$ denotes the diagonal part of $V$ and $V^{\mathrm{od}}$ the off-diagonal part. 
We will in particular consider the scaling in \eqref{eq:Vscaling} in the limit of weak coupling, $\lambda \ll 1$.
The parameter $\kappa \in \R$ regulates the relative strength between the intra-band coupling $V^{\mathrm{d}}$ and inter-band coupling $V^{\mathrm{od}}$. We point out that $\kappa$ does not have a sign, which means that the inter-band coupling can be either attractive or repulsive. To indicate the dependence on the parameters $\lambda$ and $\kappa$, we shall henceforth write $T_c = T_c(\lambda, \kappa)$. 

Similar to previous works \cite{cuenin.merz.2021, Hainzl.Seiringer.2008, hainzl.seiringer.2010, Henheik.Lauritsen.ea.2023}, in the weak coupling limit, a special role is played by the self-adjoint trace-class operator $\mathcal{V} : \bigoplus_{a=1}^n L^2(S_a) \to \bigoplus_{a=1}^n L^2(S_a)$, measuring the strength of the interaction matrix $V$ on the Fermi surfaces $S_a$ and $S_b$ whose action is defined as 
\begin{equation} \label{eq:calVdef}
	(\mathcal{V} u)_a(p) := \sum_{b=1}^n (\mathcal{V}_{ab}u_b)(p) 
  := \sum_{b=1}^n \frac{1}{(2 \pi)^{d/2}} \frac{2}{\sqrt{\abs{\nabla \epsilon_a(p)}}}\int_{S_b} \frac{\hat{V}_{ab}(p-q)}{\sqrt{\abs{\nabla \epsilon_b(q)}}} u_b(q) \D \omega(q) \,.
\end{equation}
Here, $\mathcal{V}_{ab} $ maps  $L^2(S_b) \to L^2(S_a)$ and we note that $\mathcal{V}_{ba} = \mathcal{V}_{ab}^*$. Moreover, corresponding to the decomposition of $V$ into diagonal and off-diagonal part in \eqref{eq:Vscaling}, we shall also write $\mathcal{V} = \mathcal{V}^{\mathrm{d}} + \kappa\mathcal{V}^{\mathrm{od}}$.
Finally, note that the pointwise evaluation of $\hat{V}_{ab}$ (and in particular on a nice codim--$1$ submanifold -- recall Assumption \ref{ass:dispersion}) is well-defined since we assume $V \in L^1(\R^d)$.

Our first result is that the effect of the inter-band coupling can only increase the critical temperature:

\begin{prop}[Increase of critical temperature] \label{thm.main.kappa.c}
Let $d\in \{1,2,3\}$ and let the dispersion relations $\epsilon_a$ satisfy Assumption \ref{ass:dispersion} 
and the interaction matrix $V = (V_{ab})_{a,b=1}^n$ satisfy Assumption \ref{ass:basic}.
Assume in addition that $V^{\rm od} \not\equiv 0$.

Then, for any $\lambda > 0$ there exists $\kappa_c^{\pm} \in [0,\infty)$ such that 
\begin{itemize}
\item 
For $\kappa \in [-\kappa_c^-, \kappa_c^+]$ 
we have $T_c(\lambda,\kappa) = T_c(\lambda,0)$, and 
\item 
For $\kappa \notin [-\kappa_c^-, \kappa_c^+]$
we have 
$T_c(\lambda,\kappa) > T_c(\lambda,0)$.
\end{itemize}
\end{prop}

The proof of \Cref{thm.main.kappa.c} is given in \Cref{sec:proof}.

\begin{remark}\label{rmk.repulsive}
In particular we note that 
$T_c(\lambda, \kappa) > 0$ for large $|\kappa|$ 
\emph{even if} $T_c(\lambda, 0) =0$.
As an example we can consider a system with $V_{ab} \geq 0$ for all $a,b$. 
For $\kappa \geq 0$ then all intra- and inter-band interactions are repulsive. 
At no inter-band coupling we have $T_c(\lambda, 0) = 0$, since a repulsive single-band system is never superconducting. 
However, for $\kappa$ large enough, the system becomes superconducting, 
\emph{even though also the inter-band interactions are repulsive.}
\end{remark}

For our second (main) result 
we will assume that at least one of the intra-band interactions $V_{aa}$ has an attractive part on the Fermi surface, meaning that 
\begin{equation} \label{eq:ea}
\mathfrak{e}_a := \inf \spec  \mathcal{V}_{aa}
\end{equation}
is strictly negative for at least one $a =1, ... , n$. Since the trace of $\mathcal{V}_{aa}$ can be computed as $\tr(\mathcal{V}_{aa}) = 2(2 \pi)^{-d/2} \hat{V}_{aa}(0)\int_{S_a}|\nabla \epsilon_a(p)|^{-1} \D \omega(p)$, a sufficient condition for $\mathfrak{e}_a < 0$ is that $\int V_{aa} < 0$. Finally, for every $a \in \{1,...,n\}$, we denote the ground state space of $\mathcal{V}_{aa}$ by 
\begin{equation} \label{eq:La}
\mathcal{L}_a := \mathrm{span} \left\{ u \in L^2(S_a) : (\mathcal{V}_{aa} - \mathfrak{e}_a)u = 0 \right\}\,. 
\end{equation}

We can now formulate our main result, the proof of which is given in Section \ref{sec:proof}.

\begin{thm}[Weak coupling] \label{thm:main}
Let $d\in \{1,2,3\}$ and assume that $\epsilon$ and $V$ satisfy \Cref{ass:basic,ass:dispersion}.
Assume in addition that $\int_{\R^d} \int_{\R^d} |V_{ab}(x)| |x-y|^2 |V_{a' b'}(y)| \D x \D y < \infty$ for all $a,b,a', b' \in \{1,...,n\}$
and that $\min_{a \in \{1,...,n\}} \mathfrak{e}_a< 0$.

Then we have the following: 
\begin{enumerate}[(i)]
\item There exist constants ${A}_1^{\pm} \in [0,\infty )$ such that for small $\lambda$ and $|\kappa|$ we have 
\begin{equation} \label{eq:main-degenerate}
\log \left(\frac{T_c(\lambda,\kappa )}{T_c(\lambda, 0)}\right) = {A}_1^{\sgn( \kappa)} \,  |\kappa| \lambda^{-1} + O(\kappa^2\lambda^{-1}) + O(\kappa).
\end{equation}
We have $A^{\pm}_1 > 0$ if and only if 
there exist a least two minima $\hat{a}_1, \hat{a}_2 \in \{1,...,n\}$ of $a \mapsto \mathfrak{e}_a$ 
and functions $u_{\hat{a}_i} \in \mathcal{L}_{\hat{a}_i}$ for $i=1,2$ such that the quadratic form 
$\langle u_{\hat{a}_1}, \mathcal{V}_{\hat{a}_1 \hat{a}_2} u_{\hat{a}_2} \rangle_{L^2(S_{\hat{a}_1})} 
= \langle \mathcal{V}_{\hat{a}_2 \hat{a}_1} u_{\hat{a}_1},  u_{\hat{a}_2} \rangle_{L^2(S_{\hat{a}_2})} \neq 0$ 
does not vanish. 

If $\hat{a}_1, \hat{a}_2$ are the only minima of $a \mapsto \mathfrak{e}_a$ and $\dim \mathcal{L}_{\hat{a}_i} = 1$ for at least one $i=1,2$, it holds that $A_1^+ = A_1^-$.

\item 
Suppose that the minimizer $\hat a$ of $a \mapsto \mathfrak{e}_a$ is unique. 
Then there exists a constant $A_2 \in [0,\infty) $ such that for small $\lambda$ and $|\kappa|$
\begin{equation} \label{eq:main}
	\log \left(\frac{T_c(\lambda,\kappa )}{T_c(\lambda, 0)}\right) 
		= A_2 \kappa^2 \lambda^{-1}
		+ O(\kappa^3 \lambda^{-1}) + O(\kappa^2)\,. 
\end{equation}
{We have $A_2 > 0$ if and only if 
$\mathcal{V}_{a\hat a}\big\vert_{\mathcal{L}_{\hat{a}}}  \not\equiv 0$ for some $a\ne \hat a$.
}
\end{enumerate}
\end{thm}

{We now informally interpret \Cref{thm.main.kappa.c} and \Cref{thm:main}  in the following Remark~\ref{rmk:qual}. }

\begin{remark}[Qualitative interpretation of our main results]	\label{rmk:qual}
\Cref{thm.main.kappa.c} says, in particular, that $T_c(\lambda,\kappa) \geq T_c(\lambda,0)$
with no assumptions on the coupling strengths $\lambda$ and $\kappa$. They may be order one or even large.
Thus, it may be understood as the statement:
\begin{enumerate}[(1)]
\item In a multi-band superconductor, the critical temperature increases when invoking inter-band couplings, 
	\emph{irrespective of both its attractive/repulsive nature and its strength}.
\end{enumerate}

{Part} (i) of \Cref{thm:main} describes a degenerate case of (at least) two bands giving rise to {(approximately)} the same critical temperature. 
Assuming further that these (at least) two bands couple non-trivially then $A_1^\pm>0$.
Rewriting \eqref{eq:main-degenerate} in this case, we have 
\begin{equation}
	T_c(\lambda,\kappa ) = T_c(\lambda,0 ) 
	\exp \left[ \frac{A_1^{\sgn( \kappa)}}{\lambda^2}|\kappa\lambda|\left[1 + O(\lambda) + O(\kappa)\right]\right]
  \approx   T_c(\lambda,0 )  \exp \left[ \frac{A_1^{\sgn( \kappa)}}{\lambda^2}|\kappa\lambda|\right].
\end{equation}
Thus, {part}  (i) may be understood as 
\begin{enumerate}[(1)]
\setcounter{enumi}{1}
\item In the degenerate case of (at least) two equally strong bands, this increase in the critical temperature is \emph{linear} 
for small inter-band coupling strengths $\kappa\lambda$.
\end{enumerate}
This effect has been previously observed in the physics literature \cite{bussmann2017multigap}, 
in contrast with the usual quadratic enhancement. 
Moreover, for three (or more) bands that give rise to the same critical temperature in the absence of interband couplings, the slope of the enhancement will generically depend on the sign of $\kappa$, i.e., $A_1^+ \neq A_1^-$; see e.g.~\cite{yerin2013ginzburg}.  
\vspace{3mm}

{Part} (ii) of \Cref{thm:main} describes the generic case of a unique band $\hat a$ being the strongest. 
Further, one usually also has that $\mathcal{V}_{a\hat{a}}\big\vert_{\mathcal{L}_{\hat{a}}} \not \equiv 0$ (meaning this band couples non-trivially to the rest). 
Thus, Theorem \ref{thm:main}~(ii) says that the constant $A_2$ generically takes value $A_2 > 0$. 
Hence, rewriting \eqref{eq:main} in that generic case, we have 
\begin{equation}
	T_c(\lambda,\kappa ) = T_c(\lambda,0 ) 
	\exp \left[ \frac{A_2}{\lambda^3}(\kappa\lambda)^2\left[1 + O(\lambda) + O(\kappa)\right]\right]
  \approx T_c(\lambda,0 )  \exp \left[ \frac{A_2}{\lambda^3}(\kappa\lambda)^2\right]
\end{equation}
Thus, {part} (ii) may be understood as follows:
\begin{enumerate}[(1)]
\setcounter{enumi}{2}
\item In the generic case of one dominating band, this increase in the critical temperature is \emph{quadratic} 
for small inter-band coupling strengths $\kappa\lambda$.
\end{enumerate}
This effect has been previously observed in the physics literature \cite{bussmann2004enhancements, bussmann2011isotope, bussmann2019multi}. 
In our approach, the quadratic enhancement of $T_c$ eventually stems from second order perturbation theory for the Birman--Schwinger operator 
associated with $K_{T_c} + V$ from \eqref{eq:Tc}. 
\end{remark}

\begin{remark}\label{rmk.2x2matrix}
As an illustrative simple example of where the `linear' versus `quadratic' increase arises, 
one may consider the eigenvalues of the $2\times 2$ matrices
\begin{equation*}
\begin{bmatrix}
1 & \kappa \\ \kappa & 1
\end{bmatrix},
\quad 
\lambda_{\textnormal{max}} = 1+\kappa,
\qquad 
\textnormal{and}
\qquad 
\begin{bmatrix}
1 & \kappa \\ \kappa & 0
\end{bmatrix},
\quad 
\lambda_{\textnormal{max}} = \frac{1 + \sqrt{1+4\kappa^2}}{2} = 1 + \kappa^2 + O(\kappa^4).
\end{equation*}
For the first matrix the dependence of the largest eigenvalue on $\kappa$ is linear, while for the second it is quadratic for small $\kappa$;
see also \Cref{rmk.two-band} below.
This is the underlying effect distinguishing the two different cases in \Cref{thm:main}.
\end{remark}

{We conclude this section by providing an explicit expression for the constant $A_2$ in a simple example and assuming the generic case of a unique strongest band.}

\begin{ex}[Explicit formula for $A_2$] \label{ex:expl} Assuming Sommerfeld dispersion relations\footnote{In case that all the dispersion relations are radially symmetric, the additional condition $\int_{\R^d} \int_{\R^d} |V_{ab}(x)| |x-y|^2 |V_{a' b'}(y)| \D x \D y < \infty$ from Theorem \ref{thm:main} can be relaxed, cf.~\cite[Theorem~2.1]{hainzl.seiringer.2010}.}
	\begin{equation} \label{eq:sommerfeld}
\epsilon_a(p) = \frac{p^2}{2m_a} - \mu_a
	\end{equation}
	with $m_a, \mu_a > 0$ for all $a \in \{1,...,n\}$ and radial interaction potentials $V_{ab}$, one can easily find an explicit expression for the constant $A_2$ in \eqref{eq:main}.
 
 Assume that $\hat{a}$ is the unique minimum of $a \mapsto \mathfrak{e}_a$ and $\dim \mathcal{L}_{\hat{a}} = 1$, which corresponds to $s$-wave superconductivity (for $\dim \mathcal{L}_{\hat{a}}>1$ the formulas below are similar). 
Let $e_1 \in \R^d$ be the unit vector in $1$-direction\footnote{By rotational invariance, one could have chosen any unit vector in $\R^d$. This shows that, in particular, $j_d$ is real-valued.} and denote 
\begin{equation} \label{eq:jddef}
	j_d(y) := \frac{1}{|\Sph^{d-1}|} \int_{\S^{d-1}} \ee^{\ii y p \cdot e_1} \D \omega(p) \quad \text{for} \quad y \in \R\,. 
\end{equation}
Moreover, for $a,b \in \{1,...,n\}$, let 
\begin{equation} \label{eq:vabdef}
	\mathfrak{v}_{ab} := \frac{|\Sph^{d-1}|}{(2 \pi)^d} (4 m_a m_b)^{d/4} (\mu_a \mu_b)^{\frac{d-2}{4}}\int_{\R^d} V_{ab}(x) j_d(\sqrt{2m_a\mu_a} |x|) j_d(\sqrt{2m_b\mu_b} |x|) \D x \,. 
\end{equation}
Note that $\mathfrak{v}_{\hat{a} \hat{a}} =  \mathfrak{e}_{\hat{a}} $ and $\mathfrak{v}_{aa} \in \spec \mathcal{V}_{aa}$ for all $a \in \{1,...,n\}$, 
showing that $|\mathfrak{v}_{\hat{a}\hat{a}} - \mathfrak{v}_{aa}| \gtrsim 1- \delta_{\hat{a}a}$, with $\delta_{ab}$ the Kronecker delta.

With these notations \eqref{eq:jddef}--\eqref{eq:vabdef}, in Section \ref{sec:proof}, we prove the constant $A_2$ to be given by
\begin{equation} \label{eq:Aform}
	A_2 = \sum_{\substack{a \neq \hat{a}}}
	 \frac{|\mathfrak{v}_{a\hat{a}}|^2}{|\mathfrak{v}_{\hat{a}\hat{a}}|^2 |\mathfrak{v}_{\hat{a}\hat{a}} - \mathfrak{v}_{aa}|}\,.
\end{equation}

Armed with \eqref{eq:Aform}, we find that $A_2 > 0$ if and only if $\mathfrak{v}_{\hat{a}a} \neq 0$ for some $a \neq \hat{a}$. On the one hand, as $\mathfrak{v}_{\hat{a}a}$ is essentially a Fourier transform, there surely exist (``ungeneric'') potentials $V_{\hat{a}a}$ such that $\mathfrak{v}_{\hat{a}a}$ vanishes for whole intervals of $m$'s and $\mu$'s (by compact support in the Fourier-type space). On the other hand, given $V_{\hat{a}a}$ with exponential decay at infinity, it is an elementary exercise, invoking analyticity in the $m$ and $\mu$ parameters, to show that the set of such values for which  $\mathfrak{v}_{\hat{a}a} = 0$ is isolated. Hence, $A_2 > 0$ is the generic scenario. 
\end{ex}

\begin{ex}[Explicit formula for $T_c$ for a two-band model]\label{rmk.two-band}
Consider the setting of radial interactions and Sommerfeld dispersions as in \Cref{ex:expl} above with $\dim \mcL_{a} = 1$ for $a=1,2$ 
(but not necessarily assuming that $\inf \mathfrak{e}_a < 0$). 
We further restrict to a two-band case and assume that for all $\kappa \in \R$ 
the ground state space of $\mcV = \mcV^{\rm d} + \kappa \mcV^{\rm od}$ is contained in $\mcL_1 \oplus \mcL_2$. 
(Physically this means that we have two coupled $s$-wave bands.)
Then, $T_c$ can be computed analytically in $\kappa$:  
In case that 
$\mathfrak{v}_{\rm min}(\kappa) := \frac{\mathfrak{v}_{11} + \mathfrak{v}_{22}}{2} - \sqrt{\left(\frac{\mathfrak{v}_{11} - \mathfrak{v}_{22}}{2}\right)^2 + \kappa^2 \abs{\mathfrak{v}_{12}}^2} < 0$ 
(which happens, e.g., if $\mathfrak{v}_{11}<0$ or $\mathfrak{v}_{22}<0$), we claim that 
\begin{equation}\label{eqn.twoband.infinite}
  T_{c}(\lambda,\kappa) 
    = T_0 \exp \left[\frac{1}{\lambda\mathfrak{v}_{\rm min}(\kappa) + O_\kappa(\lambda^2)}\right]
\end{equation}
for some fixed temperature scale $T_0$, and where the implicit constant in $O_\kappa(\lambda^2)$ depends on $\kappa$.
Note that this expression for the critical temperature in a two-band superconductor already appeared in the seminal paper by Suhl--Matthias--Walker \cite{Suhl.Matthias.ea.1959}. 
We point out that, even if $\mathfrak{v}_{ab} > 0$ for all $a,b=1,2$, i.e., all interactions are repulsive, 
we have $\mathfrak{v}_{\rm min}(\kappa) < 0$ for $|\kappa|$ large enough; recall also \Cref{rmk.repulsive}. 
Further, we point out that the leading-order $\kappa$-dependence in \eqref{eqn.twoband.infinite} is linear respectively quadratic
in case $\mathfrak{v}_{11} = \mathfrak{v}_{22}$ respectively $\mathfrak{v}_{11}\ne \mathfrak{v}_{22}$,
{matching the two different settings in \Cref{thm:main}}; recall also \Cref{rmk.2x2matrix}.
Finally, on the other hand, if $\mathfrak{v}_{\rm min}(\kappa) > 0$, then $T_c(\lambda,\kappa)=0$ for $\lambda$ small enough.

The formula \eqref{eqn.twoband.infinite} above follows from the proof of \Cref{thm:main}. More concretely, inspecting the proof of \Cref{thm:main} below
{and expanding the formula $-1 = \inf \spec S_{T_c(\lambda,\kappa)}$ to first order in $\lambda$}, the formula follows. 
\end{ex}

In the following Section \ref{sec:proof} we prove \Cref{thm.main.kappa.c}, \Cref{thm:main} {and the explicit formula \eqref{eq:Aform} for~$A_2$.}

\section{Proofs} \label{sec:proof}

We first give the 

\begin{proof}[{Proof of \Cref{thm.main.kappa.c}}]
Consider \eqref{eq:Tc}. At $\kappa=0$ and $T=T_c(\lambda,0)$ we thus find for any $\lambda > 0$ that 
$K_{T_c(\lambda,0)} + \lambda V^{\rm d} \geq 0$. 
Define then the function by the variational principle
\begin{equation*}
f(\kappa) = \inf \spec (K_{T_c(\lambda,0)} + \lambda V^{\rm d} + \lambda \kappa V^{\rm od})
= \inf_{\psi: \norm{\psi}_{L^2}=1} \longip{\psi}{K_{T_c(\lambda,0)} + \lambda V^{\rm d} + \lambda \kappa V^{\rm od}}{\psi}
.
\end{equation*}
Being an infimum over affine functions,  $f$ is concave in $\kappa$. 
Moreover, since $V^{\rm od}$ is 
off-diagonal $f$ assumes a local maximum at $\kappa=0$, again by the variational principle. 
It follows that $f(\kappa)\leq 0$ for all $\kappa$ and with equality only on some (possibly infinite) interval $[-\kappa_c^-, \kappa_c^+]$. 
Since $K_T$ is a monotone increasing function of $T$ then $T_c(\lambda,\kappa) \geq T_c(\lambda,0)$
with strict inequality outside the interval $[-\kappa_c^-, \kappa_c^+]$.

To see that the interval $[-\kappa_c^-, \kappa_c^+]$ is finite 
we note that, since $V^{\rm od}$ is off-diagonal, 
we can find a function $\psi\in H^1$ (of finite kinetic energy) with 
$\longip{\psi}{\lambda V^{\rm od}}{\psi} \leq -e < 0$ for some $e > 0$.
Then, by Sobolev's inequality \cite[Theorem 8.3]{analysis}, we have 
$\longip{\psi}{K_{T_c(0)} + \lambda V^{\rm d}}{\psi} \leq C$ and so, 
by the variational principle
\begin{equation}\label{eqn.f.kappa.<0}
f(\kappa) \leq \longip{\psi}{K_{T_c(\lambda,0)} + \lambda V^{\rm d} + \lambda \kappa V^{\rm od})}{\psi} \leq C - \kappa e < 0,
\end{equation}
for $\kappa > C/ e$. 
Thus $\kappa_c^+ < C/ e$. Similarly $\kappa_c^-$ is finite. 
By \eqref{eq:Tc}, this concludes the proof.
\end{proof}

Next, we give the

\begin{proof}[{Proof of \Cref{thm:main}}]
First, we note that $\kappa=0$ correspond to decoupled one-band models. Thus, $T_c(\lambda,0)>0$ by \cite{frank.hainzl.naboko.seiringer}.
In particular then by \Cref{thm.main.kappa.c} we have $T_c(\lambda,\kappa) \geq T_c(\lambda,0) > 0$.

Next, analogously to the single-band case 
\cite{frank.hainzl.naboko.seiringer, Hainzl.Seiringer.2008, hainzl.seiringer.scat.length, hainzl.seiringer.scat.length, Henheik.2022, Henheik.Lauritsen.ea.2023}
we use the Birman--Schwinger principle to 
relate spectral properties of the unbounded operator $K_T + \lambda V$ to the compact  Birman--Schwinger operator 
\begin{equation} \label{eq:BSoperator}
B_T := \lambda V^{1/2} K_T^{-1} |V|^{1/2}.
\end{equation}
In \eqref{eq:BSoperator}, we used a polar decomposition $V= U |V|$ for the self-adjoint interaction matrix $V$ and denoted $V^{1/2} := U |V|^{1/2}$. 
Note that $B_T$ has real spectrum: Indeed, $B_T$ is isospectral to\footnote{This follows from the general fact that 
$\spec(AB)\setminus\{0\} = \spec(BA)\setminus\{0\}$ for any bounded operators $A,B$. 
Moreover, in our case, $0$ is in both spectra, since $AB$ and $BA$ are compact operators on an infinite-dimensional space.} 
the self-adjoint operator 
$\lambda K_T^{-1/2} |V|^{1/2} V^{1/2} K_T^{-1/2} = \lambda K_T^{-1/2} V K_T^{-1/2}$ since $U$ and $|V|$ commute. 
The Birman-Schwinger principle then says that $-1$ is the lowest eigenvalue of $B_T$ exactly for $T=T_c$, see \cite{frank.hainzl.naboko.seiringer}.

Further, 
we decompose the Birman--Schwinger operator into a dominant singular and a bounded error term as
\begin{equation} \label{eq:BSdecomp}
B_T = \lambda \log \left(\frac{T_0}{T}\right)V^{1/2} \mathfrak{F}^\dagger  \mathfrak{F} |V|^{1/2} + \lambda V^{1/2} M_T |V|^{1/2}
\end{equation}
with all the operators in \eqref{eq:BSdecomp} being matrices and $T_0 > 0$ a fixed reference temperature. 
More precisely, we introduced the rescaled restricted Fourier transforms $\mathfrak{F} := \diag(\mathfrak{F}_{a})_{a=1}^n $ with 
\begin{equation*}
	\mathfrak{F}_{a} : L^1(\R^{d}) \to L^2(S_a) \,, \quad \text{where} \quad 
  (\mathfrak{F}_{a} \psi )(p) := \frac{1}{(2 \pi)^{d/2}} 
  \frac{\sqrt{2}}{\sqrt{|\nabla \epsilon_{a}(p)|}}
  \int_{\R^d} \ee^{- \I p \cdot x} \psi(x)\D x \bigg|_{p \in S_a}
\end{equation*}
and $M_{T} := \diag(M_{T,a})_{a=1}^n$ is such that \eqref{eq:BSdecomp} holds.

The boundedness of the second summand in \eqref{eq:BSdecomp} is the content of the following lemma, {which is rather standard in the context of BCS theory.}
We give the proof below. 
\begin{lemma}[{cf. \cite[Lemma 2]{frank.hainzl.naboko.seiringer}, \cite[Lemma 3.4]{Henheik.Lauritsen.ea.2023}, and \cite[Lemma 3.2]{hainzl.seiringer.2010}}]
\label{lem:Mbound}~\\
Under the conditions of Theorem \ref{thm:main}, we have that, uniformly in $T> 0$, $V^{1/2} M_T |V|^{1/2}$ is a bounded operator, $\sup_{T> 0}\Vert V^{1/2} M_T |V|^{1/2} \Vert \le C$. 
\end{lemma}

Using Lemma \ref{lem:Mbound}, we find, by a similar argument as in \cite{Henheik.Lauritsen.2023} (see also \cite{Henheik.Lauritsen.ea.2023,Hainzl.Seiringer.2008}), 
that 
\begin{equation*}
 S_T := \lambda \log \left(\frac{T_0}{T}\right) \mfF |V|^{1/2} \frac{1}{1 + \lambda V^{1/2} M_T |V|^{1/2}} V^{1/2} \mfF^\dagger 
\end{equation*}
has $-1$ as its lowest eigenvalue exactly for $T=T_c$.
Similarly to \cite{Henheik.Lauritsen.2023} we wish to use this fact for the two settings with or without interband coupling. 
To do this we first note that,
to leading order in $\lambda$, $S_T$ is proportional to $\mathfrak{F} V \mathfrak{F}^\dagger =  \mathcal{V}$.
Thus, the ground states of $S_T$ are among those of $\mcV$ for $\lambda$ small enough. 
(The ground states of $\mcV$ may have different $S_T$-expectations to higher order in $\lambda$.
The ground states of $S_T$ are those with smallest higher-order terms.)
As $\mcV$ depends on $\kappa$ so does the ground state space of $S_T$. Denote this space by $\mcL^\kappa$.

Next, consider degenerate perturbation theory (in $\kappa$) for the operator $S_T$. 
From degenerate first order perturbation theory for the ground states, 
we find that any (normalized) ground state $u_\kappa \in \mcL^\kappa$ of $S_{T_c(\lambda,\kappa)}$
can be written as 
\begin{equation*}
u_\kappa 
	= \frac{u^{(0)} + \kappa  u^{(1)}}{\sqrt{1 + \kappa^2 \ip{u^{(1)}}{u^{(1)}}}},
	\quad u^{(0)} \in \mcL^0,
	\quad 
	u^{(1)} \perp \mcL^0,
	\quad 
	\ip{u^{(0)}}{u^{(0)}} = 1, 
	\quad  
	\ip{u^{(1)}}{u^{(1)}} \leq C.
\end{equation*}
Then, for small enough $\lambda$, we have $-1 = \longip{u_\kappa}{S_{T_c(\lambda,\kappa)}}{u_\kappa}$ and 
$-1 = \longip{u^{(0)}}{S_{T_c(\lambda,0)}}{u^{(0)}}$. 
Next, writing $\frac{1}{1+x} = 1 - \frac{x}{1+x}$ we decompose $S_T$ as 
\begin{equation*}
S_T = \lambda \log\left(\frac{T_0}{T}	\right) \left[\mcV - \lambda \mfF |V|^{1/2} \frac{V^{1/2} M_T |V|^{1/2}}{1 + \lambda V^{1/2}M_T |V|^{1/2}} V^{1/2} \mfF^\dagger\right].	
\end{equation*}
At the critical temperature $T_c(\lambda,\kappa)$ we write the second term in $[\ldots]$ as $\lambda \mcW(\lambda,\kappa)$.
Combining then for both with and without inter-band couplings we get 
(noting that $u_\kappa$ and $u^{(0)}$ are ground states of $\mcV^{\rm d} + \kappa \mcV^{\rm od}$ and $\mcV^{\rm d}$ respectively)
\begin{equation*}
\begin{aligned}
& \lambda \log\left(\frac{T_c(\lambda,\kappa)}{T_c(\lambda,0)}\right)
	\\ &\quad  = \frac{1}{\inf\spec\left(\mcV^{\rm d} + \kappa \mcV^{\rm od}\right) 
		- \lambda \longip{u_\kappa}{\mcW(\lambda,\kappa)}{u_\kappa}}
		- \frac{1}{\inf\spec\left(\mcV^{\rm d}\right) 
		- \lambda \longip{u^{(0)}}{\mcW(\lambda,0)}{u^{(0)}}}.
\end{aligned}	
\end{equation*}
To evaluate this, we first note that, 
by simple perturbation theory for the compact self-adjoint operator 
$\mathcal{V}^{\mathrm{d}} + \kappa \mathcal{V}^{\mathrm{od}}$ up to second order 
\begin{equation} \label{eq:pertheor} 
\inf \spec \big(\mathcal{V}^{\mathrm{d}} + \kappa \mathcal{V}^{\mathrm{od}} \big) 
	= \mathfrak{e}_{\hat{a}} - U_1^{\sgn(\kappa)} |\kappa| - U_2^{\sgn(\kappa)} \kappa^2 + O(\kappa^3)
\end{equation}
for some constants $U_1^\pm, U_2^\pm \ge 0$. 
(The signs of $U_1^\pm$ follows from the the fact that $T_c(\lambda,\kappa)\geq T_c(\lambda,0)$. \normalcolor
The signs of $U_2^\pm$ is a general feature of second order perturbation theory.) 
Second, expanding $\longip{u_\kappa}{\mcW(\lambda,\kappa)}{u_\kappa}$ in powers of $\kappa$, we have by \Cref{lem:Mbound} {(and using that $\mathfrak{F}$ and $\mathfrak{F}^\dagger$ as well as multiplication by $|V|^{1/2}$ and $V^{1/2}$ are bounded operators with the appropriate (co)domains)}
\begin{equation}\label{eqn.expand.W.kappa}
\begin{aligned}
& \longip{u_\kappa}{\mcW(\lambda,\kappa)}{u_\kappa}
	\\ & \quad = \longip{u^{(0)}}{\mcW(\lambda,0)}{u^{(0)}}
		+ \kappa \left[\longip{u^{(0)}}{\partial_\kappa \mcW(\lambda,0)}{u^{(0)}} + 2\tRe \longip{u^{(0)}}{\mcW(\lambda,0)}{u^{(1)}}\right]	
		+ O(\kappa^2).
\end{aligned}
\end{equation}

To prove \Cref{thm:main}~(i) we use \eqref{eq:pertheor} and \eqref{eqn.expand.W.kappa} to order $\kappa$. 
Bounding further $\norm{\mcW(\lambda,0)} \leq C$ by \Cref{lem:Mbound} {again} it follows that 
\begin{equation*}
\lambda \log\left(\frac{T_c(\lambda,\kappa)}{T_c(\lambda,0)}\right)
	= \frac{U_1^{\sgn(\kappa)}}{\mathfrak{e}_{\hat{a}}^2} |\kappa| + O(\kappa^2) + O(|\kappa| \lambda).
\end{equation*}
To prove the second part of \Cref{thm:main}~(i) we note that 
$A_1^\pm = U_1^\pm/\mathfrak{e}_{\hat a}^2 > 0$ \normalcolor happens when first order perturbation theory in \eqref{eq:pertheor} does not vanish. 
This is the case, precisely if there exist a least two minima $\hat{a}_1, \hat{a}_2 \in \{1,...,n\}$ 
of $a \mapsto \mathfrak{e}_a$ and functions $u_{\hat{a}_i} \in \mathcal{L}_{\hat{a}_i}$ for $i=1,2$ 
such that the quadratic form 
$\langle u_{\hat{a}_1}, \mathcal{V}_{\hat{a}_1 \hat{a}_2} u_{\hat{a}_2} \rangle_{L^2(S_{\hat{a}_1})} 
= \langle \mathcal{V}_{\hat{a}_2 \hat{a}_1} u_{\hat{a}_1},  u_{\hat{a}_2} \rangle_{L^2(S_{\hat{a}_2})} \neq 0$ 
does not vanish. Finally, one can easily check that, in case of exactly two minima $\hat{a}_1, \hat{a}_2$ of $a \mapsto \mathfrak{e}_a$ and $\dim \mathcal{L}_{a_i} = 1$ for at least one $i=1,2$, the constants $U^\pm_1$ in \eqref{eq:pertheor} and hence $A_1^\pm$agree.

Next, to prove \Cref{thm:main}~(ii) we use \eqref{eq:pertheor} and \eqref{eqn.expand.W.kappa} to order $\kappa^2$.
By assumption, the order $\kappa$ term in \eqref{eq:pertheor} vanishes using the argument from above. 
Further, in this case also $U_2 := U_2^+ = U_2^-$ agree.
We claim that  also the order $\kappa$ terms in \eqref{eqn.expand.W.kappa} vanish. 
This follows from the fact that $\mcW(\lambda,0)$ is diagonal and the perturbation is off-diagonal.
More precisely, since $\partial_\kappa \mcW(\lambda,0)$ is off-diagonal 
we conclude that $\longip{u^{(0)}}{\partial_\kappa \mcW(\lambda,0)}{u^{(0)}} = 0$. 
Second, $\mcW(\lambda,0)u^{(0)} \in \mcL^0$ since $\mcL^0$ is a subset of the ground state space of $\mcV$ 
and is an eigenspace for $S_T$. Thus, $\longip{u^{(0)}}{\mcW(\lambda,0)}{u^{(1)}} = 0$.
{Combining $\norm{\mcW(\lambda,0)} \leq C $ (by Lemma \ref{lem:Mbound} again) with \eqref{eq:pertheor}, we conclude that }
\begin{equation*}
\lambda \log\left(\frac{T_c(\lambda,\kappa)}{T_c(\lambda,0)}\right)
	= 
	\frac{U_2}{\mathfrak{e}_{\hat{a}}^2} \kappa^2
	+ O(\kappa^3) + O(\lambda\kappa^2) \,,
\end{equation*}
which immediately shows the first part of Theorem \ref{thm:main}~(ii). 
Lastly, $A_2 = U_2/\mathfrak{e}_{\hat a}^2 = 0$ happens when both first and second order perturbation theory vanish in \eqref{eq:pertheor}. 
This is precisely the case if 
$\mathcal{V}_{a\hat{a}}\big\vert_{\mathcal{L}_{\hat{a}}}  \equiv 0$ for all $a \neq \hat{a}$.

This concludes the proof of Theorem \ref{thm:main}. 
\end{proof}

It remains to give the

\begin{proof}[{Proof of \Cref{lem:Mbound}}] 
The argument is very similar to \cite[Lemma~3.2]{hainzl.seiringer.2010} for dimensions $d=2,3$ (the adjustments to $d=1$ are straightforward), hence we will be very brief. 
	
First, multiplying by the unitary $U^*$, we see that $ |V|^{1/2} M_T |V|^{1/2}$ is self-adjoint and satisfies $\Vert |V|^{1/2} M_T |V|^{1/2} \Vert = \Vert V^{1/2} M_T |V|^{1/2} \Vert$. Then, for $\psi \in L^2(\R^d, \C^n)$ setting $\varphi := |V|^{1/2} \psi$, it holds that 
\begin{equation} \label{eq:Mboundstart}
\longip{\psi}{|V|^{1/2} M_T |V|^{1/2}}{\psi} = \sum_{a=1}^n \left[ \int_{\R^d} \frac{|\hat{\varphi}_a(p)|^2}{K_{T,a}(p)} \D p-2 \log\left(\frac{T_0}{T}\right) \int_{S_a} \frac{|\hat{\varphi}_a(p)|^2}{|\nabla \epsilon_a(p)|} \D \omega(p)\right]
\end{equation}
since $\mathfrak{F}^\dagger \mathfrak{F}$ and $K_T^{-1}$ are diagonal, and we denoted $K_{T,a}(p) := \epsilon_{a}(p)/\tanh\left(\frac{\epsilon_a(p)}{2T}\right)$. 
The expression in \eqref{eq:Mboundstart} is the analog of \cite[Eq.~(3.25)]{hainzl.seiringer.2010} with the identifications 
$T(p) + e \to (K_{T,a}(p)- 2T) + 2T$, $f(e) \to 2 \log(T_0/T)$, and $\nabla P \to \nabla \epsilon_a$. 
Now, we estimate every summand in \eqref{eq:Mboundstart} separately, 
following the arguments in \cite[pp.~496--498]{hainzl.seiringer.2010} for $r=1$\footnote{Note that the second to last line in \cite[p.~497]{hainzl.seiringer.2010} contains a misprint: The estimate $\int_{0}^{\tau} t(t^2 + e)^{-1} \D t \le \mathrm{const} \, g(e)$ should rather be $\int_{0}^{\tau} t(t^r + e)^{-1} \D t \le \mathrm{const} \, g(e)$.} 
with the following modifications:
The bound \cite[Eq.~(3.33)]{hainzl.seiringer.2010} is replaced by 
\begin{equation*}
	\begin{split}
|\hat{\varphi}_a(p)|^2 &\lesssim \sum_{b ,b' = 1}^n\iint (|V|^{1/2})_{ab}(x) (|V|^{1/2})_{ab'}(y) \overline{\psi_b(x)} \psi_{b'}(y) \mathrm{e}^{\ii p \cdot (x-y)} \D x \D y \\
&\lesssim \max_{a,b=1}^{n} \Vert V_{ab} \Vert_{L^1(\R^d)} \Vert \psi \Vert_{L^2(\R^d, \C^n)}^2 \lesssim \Vert \psi \Vert_{L^2(\R^d, \C^n)}^2
	\end{split}
\end{equation*}
and, in a similar fashion, the bound \cite[Eq.~(3.34)]{hainzl.seiringer.2010} is replaced by
\begin{equation*}
	\begin{split}
\big| \nabla|\hat{\varphi}_a(p)|^2 \big| &\lesssim \left(\max_{b, b'=1}^n \iint |V|_{ab}(x) |x-y|^2 |V|_{ab'}(y) \D x \D y \right) \Vert \psi \Vert_{L^2(\R^d, \C^n)}^2  \\ 
&\lesssim \left(\max_{a,b, a', b'=1}^n \iint |V_{ab}|(x) |x-y|^2 |V_{a'b'}|(y) \D x \D y \right) \Vert \psi \Vert_{L^2(\R^d, \C^n)}^2 \lesssim \Vert \psi \Vert_{L^2(\R^d, \C^n)}^2\,. 
	\end{split}
\end{equation*}
In both estimates, the second step follows from elementary linear algebra. This concludes the proof of Lemma \ref{lem:Mbound}. 
\end{proof}

Finally, we give the

\begin{proof}[{Proof of \Cref{eq:Aform}}]
From the proof of \Cref{thm:main} we have $ A_2 = U_2/\mathfrak{e}_{\hat{a}}^2$.

Since $V_{aa}$ is radial by assumption, all eigenfunctions of $\mathcal{V}_{aa}$ are given by appropriately normalized $d$-dimensional spherical harmonics $(2 m_{a}\mu_{a})^{-(d-1)/4}Y_\ell$ and their eigenvalues are denoted $\mathfrak{e}_a(\ell)$. Here, we abused the common notation $Y$ from three spatial dimensions for all $d=1,2,3$ and regard $\ell$ as an angular momentum multi-index. The $Y_\ell$ form an orthonormal basis of $L^2(\Sph^{d-1})$, i.e.~$\langle Y_\ell, Y_{\ell'} \rangle_{L^2(\Sph^{d-1})} = \delta_{\ell \ell'}$, and $Y_0$ is always understood to be the constant spherical harmonic. Moreover, we shall also write $u_a(\ell) := (0,...,0, (2 m_{a}\mu_{a})^{-(d-1)/4}Y_\ell, 0,...,0)  \in \bigoplus_{a=1}^n L^2(S_a)$, where the only non-zero entry is at position $a \in \{1,...,n\}$.

With these notations, using $\dim \mathcal{L}_{\hat{a}} = 1$, it follows that $u_{\hat{a}}(0)$ is the unique (normalized) ground state vector of $\mathcal{V}^{\mathrm{d}}$. 
Hence, second order perturbation theory shows that $U_2$ can be evaluated as
\begin{equation} \label{eq:Aexpl1}
U_2 = \sum_{(a, \ell) \neq (\hat{a}, 0)} \frac{|\langle u_{a}(\ell), \mathcal{V}^{\mathrm{od}} u_{\hat{a}}(0) \rangle|^2}{\mathfrak{e}_a(\ell) - \mathfrak{e}_{\hat{a}}} \,. 
\end{equation}
Since all the $V_{ab}$ are radial by assumption and using orthogonality of spherical harmonics, the large sum in \eqref{eq:Aexpl1} (over all bands and angular momenta) further collapses to a sum only over the bands, i.e.
\begin{equation*}
U_2 = \sum_{a \neq \hat{a}} \frac{|\langle u_{a}(0), \mathcal{V}^{\mathrm{od}} u_{\hat{a}}(0)\rangle |^2}{|\mathfrak{e}_a(0) - \mathfrak{e}_{\hat{a}}|} \,,
\end{equation*}
where we put the absolute value in the denominator for better comparability with \eqref{eq:Aform}. Indeed, in order to arrive at \eqref{eq:Aform}, we simply note that $\mathfrak{e}_{\hat{a}} = \mathfrak{v}_{\hat{a}\hat{a}}$, $\mathfrak{e}_a(0) = \mathfrak{v}_{aa}$, as well as $ \langle u_{a}(0), \mathcal{V}^{\mathrm{od}} u_{\hat{a}}(0)\rangle = \mathfrak{v}_{a\hat{a}}$, where $\mathfrak{v}_{ab}$ was defined in \eqref{eq:vabdef}. 
\end{proof}

\bigskip

\noindent \textit{Acknowledgments.}
We would like to thank J. Lenells and R. Seiringer for their interest and helpful discussions, and E. Babaev and Y. Yerin for useful comments about the literature.
J.H.~gratefully acknowledges partial financial support by the ERC Advanced Grant ``RMTBeyond'' No.~101020331. {E.L.  gratefully acknowledges support from the Swedish Research Council, Grant No.\ 2023-04726}. 
A.B.L. gratefully acknowledges partial financial support by the Austrian Science Fund (FWF) through 
grant \href{https://doi.org/10.55776/I6427}{DOI: \nolinkurl{10.55776/I6427}}
(as part of the SFB/TRR~352).

\bigskip

\noindent \textit{Data Availability.} Data sharing is not applicable to this article as no new data were created or analyzed in this study.

\appendix
\section{Derivation of the multi-band BCS functional}\label{app:derivation}
We give here a heuristic derivation of the functional in \eqref{eq:mbBCS}, see also \cite{leggett.diatomic} and \cite[Appendix~A]{hainzl.hamza.seiringer.solovej}.
It arises as a formal infinite volume limit of the (negative) pressure functional restricted to a certain class of states.

To start, we consider a spin-$\frac{1}{2}$ Fermi gas localized to some box $\Lambda = [0,L]^d$ 
{so that Fourier space is $\Lambda^*=\frac{2\pi}{L}\Z^d$ (we use periodic boundary conditions for the fermion operators).}
The fermions come in $n$ different species --- these are the \emph{bands}. 
We consider the most general interactions carrying $4$ band indices.   
Thus, the Hamiltonian is given by 
\begin{equation*}
H = \sum_{k,\sigma}\sum_{a} \epsilon_{a}^{(0)}(k) a_{k,\sigma,a}^* a_{k,\sigma,a} 
  + \frac{1}{2L^d} \sum_{p,k,k',\sigma,\tau} \sum_{a,a',b,b'} \hat V_{aa',bb'}(p) a_{k+p,\sigma,a}^* a_{k'-p,\tau,a'}^* a_{k',\tau,b'} a_{k,\sigma,b}
\end{equation*}
with {(bare)} dispersion relations $\epsilon^{(0)}_a$ and interactions $V_{aa',bb'}$ {(we abuse notation slightly and use the same symbol $a$ both for fermin operators and a band index).} 
Here $a_{k,\sigma,a}^*$ and $a_{k,\sigma,a}$  denote the creation and annihilation operators of a particle of momentum {$k\in\Lambda^*$}, spin $\sigma{\in\{\uparrow,\downarrow\}}$ and in band $a$. 
Next, we restrict $H$ to quasi-free states with no fixed particle number. 
These are the states relevant for BCS theory \cite{hainzl.hamza.seiringer.solovej,leggett.diatomic,bcs.original}.
Quasi-free states obey the Wick rule 
\begin{equation*}
\expect{a_1^\#a_2^\#a_3^\# a_4^\#}
  = \expect{a_1^\# a_4^\#} \expect{a_2^\#a_3^\#}
    -\expect{a_1^\#a_3^\# } \expect{a_2^\#a_4^\#}
    + \expect{a_1^\#a_2^\#}\expect{a_3^\# a_4^\#},
\end{equation*}
with each $a_i^\#$ being either $a_{k,\sigma,a}$ or $a_{k,\sigma,a}^*$. The three terms are the direct, exchange and pairing term respectively.
Then, the expectation $\expect{H}$ can be written in terms of 
\begin{equation*}
\hat \gamma_{(a,\sigma), (b,\tau)}(k,k') = \expect{a_{k,\sigma,a}^* a_{k',\tau,b}} 
\quad \text{and}\quad 
\hat \alpha_{(a,\sigma), (b,\tau)}(k,k') = 
{\expect{a_{k',\tau,b}a_{k,\sigma,a} }};
\end{equation*}
{note that the functions $\alpha_{(a,\sigma), (b,\tau)}(k,k')$ are odd under the exchange $(k,\sigma,a)\leftrightarrow (k',\tau,b)$.}
Next, we make the following three simplifying assumptions:
\begin{enumerate}[(a)]
\item We consider only translation invariant states. This means that for the one-particle density matrix $\gamma$ that $k=k'$ for the non-zero entries
and for the pairing function $\alpha$ that $k=-k'$ for the non-zero entries. 

\item 
{We consider only spin rotation invariant states  and, in particular, only spin singlet superconducting states. 
(In the mathematical physics literature, this is referred to as $SU(2)$-invariance.)}
\item 
We consider only pairing functions with one band index, meaning that for $\alpha_{(a,\sigma), (b,\tau)}$ only terms with $a=b$ are non-zero. 

\end{enumerate}
Points (a) and (b) are discussed at length in \cite{hainzl.seiringer.16}{; the simplification in  (c) is common in the multi-band physics literature \cite{Suhl.Matthias.ea.1959, moskalenko1959superconductivity, kondo1963superconductivity, leggett1966number}.}

With these assumptions, the expectation $\expect{H}$ can be written in terms of the simpler functions
\begin{equation*}
\hat \gamma_{ab}(k) = \expect{a_{k,\uparrow,a}^* a_{k,\uparrow,b}} = \expect{a_{k,\downarrow,a}^* a_{k,\downarrow,b}} \quad \text{and}\quad 
\hat \alpha_{a}(k) = \expect{a_{-k,\uparrow,a} a_{k,\downarrow,a}} = - \expect{a_{-k,\downarrow,a} a_{k,\uparrow,a}}; 
\end{equation*}
{note that the latter condition means that we only consider spin-singlet superconducting states, and this condition and the canonical anti-commutator relations imply that $\hat\alpha_a(k)$ are even functions of $k$.}
The matrix-valued functions $\hat \gamma = [\hat \gamma_{ab}]_{a,b = 1}^n$ and $\hat \alpha = \diag[\hat \alpha_a]_{a=1}^n$ 
can be conveniently grouped together in the generalized reduced one-particle density matrix $\Gamma$ given by 
\begin{equation*}
\hat \Gamma(p) = \begin{bmatrix}
\hat \gamma(p) & \hat \alpha(p) \\ \overline{\hat \alpha(p)} & \mathbbm{1} - \hat \gamma(-p)
\end{bmatrix} 
\end{equation*}
As a matrix this satisfies $0\leq \hat \Gamma(p) \leq \mathbbm{1}$ pointwise.
The expectation of $H$ in the state with generalized reduced one-particle density matrix $\Gamma$ is then
\begin{equation*}
\begin{aligned}
\expect{H}_\Gamma
	& = 2\sum_{k,a} \epsilon_a(k)^{(0)} \hat \gamma_{aa}(k)
		+ \frac{2}{L^d} \sum_{a,a',b,b'} \hat V_{aa',bb'}(0) \left[\sum_k \hat \gamma_{ab}(k)\right] \left[\sum_k \hat \gamma_{a'b'}(k)\right]
	\\ & \quad 
		- \frac{1}{L^d} \sum_{k,p,a,a',b,b'} \hat V_{aa'bb'}(p) \hat \gamma_{ab'}(p-k) \hat \gamma_{a'b}(p)
		+ \frac{1}{L^d} \sum_{k,p,a,b} \hat V_{aa,bb}(p) \overline{\hat \alpha_{a}(k-p)}  \hat \alpha_{b}(k)
	\\ & = 2\sum_{k,a} \epsilon_a(k)^{(0)} \hat \gamma_{aa}(k)
		+ \iint \sum_{a,b} V_{aa,bb}(x-y) \overline{\alpha_{a}(x-y)} \alpha_{b}(x-y) \ud x \ud y\,, 
	\\ & \quad 
		+ \iint \sum_{a,a',b,b'} V_{aa',bb'}(x-y) \left[2\gamma_{ab}(0)\overline{\gamma_{a'b'}(0)} -  \gamma_{a'b}(x-y)\overline{\gamma_{ab'}(x-y)}\right] \ud x \ud y
\end{aligned}
\end{equation*}
where the factors of two arise from the spin degrees of freedom. 
The pressure functional of the state (with generalized reduced one-particle density matrix) $\Gamma$ is then {given by}
\begin{equation*}
\mcP[\Gamma] = \frac{1}{L^d}\left[2T S(\Gamma) - \expect{H}_\Gamma\right], 
\qquad S(\Gamma) = - \sum_p \Tr_{\C^{2n}} \left[\hat\Gamma(p) \ln \hat \Gamma(p)\right]\,, 
\end{equation*}
where the factor two in front of the entropy again comes from the spin degrees of freedom.
Taking a formal infinite volume limit of $-\frac{1}{2} \mcP$ and replacing $V \to 2V$ we find the functional 
\begin{equation*}
\begin{aligned}
\mcF^{(0)}[\Gamma]
	& 
	= \int \sum_a \epsilon_a^{(0)}(p) \hat \gamma_{aa}(p) \ud p 
		- T S[\Gamma]
		+ \int_{\R^d} \sum_{a,b} \overline{\alpha_{a}(x)} V_{aa,bb}(x) \alpha_{b}(x) \ud x 
	\\ & \quad 
		+ \int_{\R^d} \sum_{a,a',b,b'} V_{aa',bb'}(x) \left[2\gamma_{ab}(0)\overline{\gamma_{a'b'}(0)} -  \gamma_{a'b}(x)\overline{\gamma_{ab'}(x)}\right] {\D x},
\end{aligned}
\end{equation*}
with the entropy per unit volume $S[\Gamma]$ as in \eqref{eqn.entropy.unit.vol.def}. 
The claim is then that, up to corrections which are negligible, the direct and exchange terms effectively only serve to renormalize the dispersions $\epsilon_a^{(0)}$, 
see \cite{leggett.diatomic}. 
Since in the interaction only terms with $a=a'$ and $b=b'$ appear,  we define the matrix $V_{ab} = V_{aa,bb}$.
Finally, by concavity of the entropy, the minimizer has $\gamma$ diagonal, 
and we may thus restrict to diagonal matrices $\gamma$.
We then recover the functional in \eqref{eq:mbBCS}.

\begin{rmk}[Brillouin zones]\label{rem:BZ}
It would be natural to consider also models with other Brillouin zones, meaning that the $k$-integrals are not over $\R^d$ but over some different domain.
For instance, a Brillouin zone of a (cubic) torus of sidelengths $2\pi/a$ corresponds to a model of fermions on a lattice $a\Z^d$ where $a>0$ is the lattice spacing.

It would be interesting to extend our results to such lattice fermion models with compact Brillouin zones, but this is beyond the scope of the present paper. 
We expect that the main results stated in the beginning of the introduction are true also for such models,
and we believe that the methods of this paper would be applicable also to these other settings.
\end{rmk}

\section{Notation}\label{sec.notation}
For the convenience of the reader, we collect some notations used throughout this paper. 

\begin{itemize} 	
\item For $f(x)$ a (suitable) function of $x=(x_1,\ldots,x_d)\in\R^d$, 
we denote its Fourier transform by $\hat f(p) = (2\pi)^{-d/2} \int_{\R^d}f(x)\mathrm{e}^{-\mathrm{i}p\cdot x}\mathrm{d}x$, (here $p\in\R^d$).
\item For $\epsilon_a(p)$ a dispersion relation determining a Fermi surface $S_a$ as explained in \Cref{ass:dispersion}, 
we denote as $\D\omega$ the (Lebesgue) measure on $S_a$. In particular, letting $\delta$ be Dirac delta, for any function $f$ the Lebesgue measure satisfies
$$
\int_{S_a} \frac1{|\nabla\epsilon_a(p)|}f(p)\D\omega(p) = \int_{\R^d}\delta(\epsilon_a(p))f(p)\D p \,. 
$$

\item The unit sphere is denoted $\S^{d-1}=\{p\in\R^d: |p|=1\}$ and the (Lebesgue) measure on $\S^{d-1}$ is denoted $\D \omega$. In particular, 
$\int_{\S^{d-1}} f(p)\mathrm{d}\omega(p) = \int_{\R^{d}}f(p)\delta(|p|-1)\mathrm{d}p$ for any $f$ and 
$|\Sph^{d-1}| = \int_{\R^{d}}\delta(|p|-1)\mathrm{d}p$.

\item For a general measure $\mu$ on $\R^d$ we denote by $L^p(\R^d,\ud\mu)$ the space of all $\C$-valued functions for which $\int |f|^p \ud \mu < \infty$. 
For Lebesgue measure $\mu$ we simply write $L^p(\R^d,\D x) = L^p(\R^d)$.

\item The space $H^{1}_{\textnormal{sym}}$ of reflection-symmetric $H^1$ functions is more concretely given as  
\begin{equation*}
H^1_{\textnormal{sym}}(\R^d,\D x) = \left\{f \in L^2(\R^d,\D x) : \hat f \in L^2(\R^d, (1+p^2)\D p), \quad f(x) = f(-x) \ \forall x \right\}.
\end{equation*}

\end{itemize}

\renewcommand*{\bibfont}{\footnotesize}
\printbibliography
\end{document}